\documentclass{article}
\usepackage{amsfonts}
\usepackage{amsmath}

\newtheorem{theorem}{Theorem}
\newtheorem{acknowledgement}[theorem]{Acknowledgement}

\input{tcilatex}

\begin{document}

\title{Schwarzschild Solution of the Generally Covariant Quaternionic Field
Equations of Sachs }
\author{Horace W. Crater\thanks{%
hcrater@utsi.edu}~, Jesse Labello, and Steve Rubenstein \\
The University of Tennessee Space Institute}
\maketitle

\begin{abstract}
Sachs has derived quaternion field equations that fully exploit the
underlying symmetry of the principle of general relativity, one in which the
fundamental 10 component metric field is replaced by a 16 component
four-vector quaternion. \ Instead of the 10 field equations of Einstein's
tensor formulation, these equations are 16 in number corresponding to the 16
analytic parametric functions $\partial x^{\mu ^{\prime }}/\partial x^{\nu }$
of the Einstein Lie Group. \ The difference from the Einstein equations is
that these equations are not covariant with respect to reflections in
space-time, as a consequence of their underlying quaternionic structure. \
These equations can be combined into a part that is even and a part that is
odd with respect to spatial or temporal reflections.\ \ This paper
constructs a four-vector quaternion solution of the quaternionic field
equation of Sachs that corresponds to a spherically symmetric static metric.
\ We show that the equations for this four-vector quaternion corresponding
to a vacuum solution lead to differential equations that are identical to
the corresponding Schwarzschild equations for the metric tensor components.
\ 
\end{abstract}

\section{Introduction}

This paper develops a solution of the quaternionic metrical field equations
of Sachs\cite{s1}, \cite{s2} ,\cite{s3} corresponding to the Schwarzschild
solution in ordinary general relativity. \ In analogy to Dirac's idea of
taking the matrix square root of the Klein Gordon equation and with it
important predictions for the electron, Sachs used quaternions to take the
square root of the metric condition\cite{s4} 
\begin{equation}
d\tau ^{2}=-g^{\mu \nu }dx_{\mu }dx_{\nu }.  \label{metric}
\end{equation}%
\ In a similar manner, Sachs uses quaternions to factorize the Einstein
equation, in effect, taking its matrix square root. We begin with a review
of the key assumptions and intermediate steps he took in deriving his
quaternionic metrical field equations before presenting our new result. \
Our review is taken from his books and key articles. \ In Sec. 1, we review
the connection between what he calls the Einstein group and the role of
quaternions. \ In Sec. 2 we give the connection between the metric and
four-vector quaternion functions. \ In Sect. 3 the derivation of the spin
affine connection is reviewed and in Sec. 4 the connections between the spin
curvature tensor and Riemann curvature tensors are established. \ \ Then in
Sec. 5 Sachs' quaternion field equation is developed. \ In Sec. 6 we
construct a four-vector quaternion that corresponds to a spherically
symmetric static metric. \ We then show that the equations for this
quaternion corresponding to a vacuum solution lead to differential equations
that are identical to the corresponding Schwarzschild equations for the
metric tensor components.

\section{The Einstein Group and Quaternions}

As with Einstein's original form of the general theory of relativity, Sachs'
metrical field equations are based on the fundamental axiom of the principle
of relativity which is 'general covariance'. \ General covariance is the
assertion that all laws of nature must be independent of the frame of
reference in which they may be represented. \ \ Sachs' 'Einstein group'
refers to the group of all analytic transformations between the space and
time coordinates of all possible frames of reference\cite{s5}. \ The
space-time transformations of the Einstein group are characterized by the
set of continuously distributed derivatives $\partial x^{\mu ^{\prime
}}/\partial x^{\nu }=x^{\mu ^{\prime }},_{\nu }$. These 16 parametric
functions are the rate of change of the space-time coordinates in one frame, 
$x^{\mu ^{\prime }},$ with respect to those of another, $x^{\nu },$ where $%
\mu ,\nu =0,1,2,3$ are the \ temporal and three spatial coordinates. \ He
states that the significance of this number is that there must be 16
independent field equations to prescribe the space-time. \ He introduces
four vector quaternion functions $q^{\mu }(x),$ with each of the four vector
components being a quaternion rather than a real number field to embody the
16 component metrical field\cite{s5}. \ He then argues that the
corresponding independent field equations should also be 16 in number.

The Einstein group is a symmetry group of general relativity and is defined
as the set of proper transformations that leave invariant the metric
condition Eq. (\ref{metric}), excluding time reversal and parity inversion.
\ This set of continuous and analytic transformations also preserves the
forms of the laws of nature. Sachs asks the question why the Einstein
equations are 10 in number rather than 16. \ His answer is that the form of
these equations are more symmetric than they need be in accordance with the
16 parameter Einstein group. \ They are not only covariant with respect to
continuous transformation, but are also covariant with respect to discrete
reflections in space and time. \ The latter is not an absolute requirement.
\ \ Sachs demonstrated with the use of the four-vector quaternion functions $%
q^{\mu }(x),$ how, in effect, the Einstein equation can factorize into two
equations, neither of which by itself is reflection symmetric or
antisymmetric\cite{s6}. \ How does this come about?

First \ one recalls that the irreducible representations of the proper
Poincaire group of special relativity obey the algebra of quaternions. \
(Related to this is the well known fact that one cannot produce parity
inversions, reflections, or time reversal by using the Pauli matrices as
generators). \ Sachs points out that the irreducible representations of the
Einstein group of general relativity also obey the algebra of quaternions.

\ Let us recall some elementary properties of quaternions. \ Recall that
Hamilton\cite{s8} introduced \ them as generalizations of complex numbers
from a two dimensional space to a four dimensional space. \ 

\begin{eqnarray}
Q &=&1x^{4}+\mathfrak{i}x^{1}+\mathfrak{j}x^{2}+\mathfrak{k}x^{3},  \notag \\
&&x^{4},x^{1},x^{2},x^{3}\text{ real,}
\end{eqnarray}%
Their conjugates are%
\begin{equation}
\bar{Q}=1x^{4}-\mathfrak{i}x^{1}-\mathfrak{j}x^{2}-\mathfrak{k}x^{3},
\end{equation}%
so%
\begin{eqnarray}
\mathfrak{i} &\mathfrak{=}&\mathfrak{-\bar{\imath}}  \notag \\
\mathfrak{j} &\mathfrak{=}&\mathfrak{-\bar{j}},  \notag \\
\mathfrak{k} &\mathfrak{=}&\mathfrak{-\bar{k}}.
\end{eqnarray}%
and requiring 
\begin{equation}
Q\bar{Q}=\left( x^{4}\right) ^{2}+\left( x^{1}\right) ^{2}+\left(
x^{2}\right) ^{2}+\left( x^{3}\right) ^{2},
\end{equation}%
implies%
\begin{eqnarray}
\mathfrak{i}^{2} &=&-1,  \notag \\
\mathfrak{j}^{2} &=&-1,  \notag \\
\mathfrak{k}^{2} &=&-1,
\end{eqnarray}%
and%
\begin{eqnarray}
\mathfrak{ij}\mathfrak{=-ji,} &&  \notag \\
\mathfrak{jk}\mathfrak{=-kj,} &&  \notag \\
\mathfrak{ki}\mathfrak{=-ik.} &&
\end{eqnarray}%
Closure implies%
\begin{eqnarray}
\mathfrak{ij} &=&\mathfrak{k=-ji,}  \notag \\
\mathfrak{ki}\mathfrak{=j} &=&-\mathfrak{ik,}  \notag \\
\mathfrak{jk} &\mathfrak{=}&\mathfrak{i=-kj.}
\end{eqnarray}

Since Pauli matrices satisfy%
\begin{equation}
\sigma _{i}\sigma _{j}=\delta _{ij}\sigma _{0}+i\varepsilon _{ijk}\sigma
_{k},
\end{equation}%
they can be used to represent quaternions if one chooses%
\begin{eqnarray}
1 &=&\sigma _{0},  \notag \\
\mathfrak{i} &\mathfrak{=}&\mathfrak{-}i\sigma _{1},  \notag \\
\mathfrak{j} &\mathfrak{=}&\mathfrak{-}i\sigma _{2},  \notag \\
\mathfrak{k} &\mathfrak{=}&\mathfrak{-}i\sigma _{3}.
\end{eqnarray}%
In compact form 
\begin{eqnarray}
Q &=&-i\sigma _{\mu }x^{\mu }=(\sigma _{0}x^{4}-i\mathbf{\sigma \cdot r),} 
\notag \\
-i\sigma _{4} &=&\sigma _{0}=%
\begin{bmatrix}
1 & 0 \\ 
0 & 1%
\end{bmatrix}%
,
\end{eqnarray}%
with a \textit{space conjugate} form%
\begin{equation}
\bar{Q}=\sigma _{0}x^{4}+i\mathbf{\sigma \cdot r,}
\end{equation}%
and \ 
\begin{equation}
Q\bar{Q}=(x^{4})^{2}+\mathbf{r}^{2}.
\end{equation}

Hamilton, of course, had no empirical reason to choose%
\begin{equation}
x^{4}=-ix^{0}=-ict.
\end{equation}%
With that choice, motivated of course by special relativity, 
\begin{equation}
Q=(\sigma _{0}x^{4}-i\mathbf{\sigma \cdot r)=-}i(\sigma _{0}x^{0}+\mathbf{%
\sigma \cdot r)=-}i\sigma _{\mu }x^{\mu },
\end{equation}%
and the invariant metric is\footnote{%
Our Minkowski metric is $\eta _{00}=-1,~\eta _{11}=\eta _{22}=\eta _{33}=1$ ,%
$\eta _{\mu \nu }=0,\mu \neq \nu $.} 
\begin{equation}
\left( x^{0}\right) ^{2}-r^{2}=-\bar{Q}Q=-x^{2}=-\eta _{\mu \nu }x^{\mu
}x^{\nu },
\end{equation}%
where%
\begin{equation}
Q=-ix^{\mu }\sigma _{\mu }=-i~%
\begin{bmatrix}
x^{0}+x^{3} & x^{1}-ix^{2} \\ 
x^{1}+ix^{2} & x^{0}-x^{3}%
\end{bmatrix}%
.
\end{equation}

Another way of writing this quaternion is to introduce the \textit{time
conjugate} operation%
\begin{eqnarray}
Q &\rightarrow &\tilde{Q}=\varepsilon Q^{\ast }\varepsilon ,  \notag \\
\varepsilon &=&i\sigma _{2}=%
\begin{bmatrix}
0 & 1 \\ 
-1 & 0%
\end{bmatrix}%
.  \label{cj}
\end{eqnarray}%
In that case%
\begin{eqnarray}
\tilde{Q} &=&ix^{\mu }\tilde{\sigma}_{\mu },  \notag \\
\tilde{\sigma}_{0} &=&-\sigma _{0},  \notag \\
\tilde{\sigma}_{i} &=&\sigma _{i}.
\end{eqnarray}%
Then%
\begin{equation}
\tilde{Q}Q=x^{\nu }x^{\mu }\tilde{\sigma}_{\nu }\sigma _{\mu }=x^{\nu
}x^{\mu }\eta _{\mu \nu }.
\end{equation}%
The simplest quaternionic four vector is the set of four constant matrices 
\begin{eqnarray}
q_{\mu } &=&\sigma _{\mu },\text{ }\mu =0,1,2,3.  \notag \\
\tilde{q}_{\mu } &=&\tilde{\sigma}_{\mu }.
\end{eqnarray}

The Lorentz transformation matrix coefficients $\alpha _{\kappa ^{\prime
}}^{\mu }$ are restricted by 
\begin{equation}
\eta _{\mu \nu }\alpha _{\kappa ^{\prime }}^{\mu }\alpha _{\lambda ^{\prime
}}^{\nu }=\eta _{\kappa ^{\prime }\lambda ^{\prime }}=\eta _{\kappa \lambda
},  \label{eta}
\end{equation}%
which gives 10 conditions on otherwise 16 independent spacetime independent
elements $\alpha _{\kappa ^{\prime }}^{\mu }$, so that, including four
space-time translations the Poincaire' group has just 6+4=10 independents
elements (parametrized additionally by three Euler angles and three boost
velocities). \ Thus%
\begin{equation}
d\tau ^{2}=-\eta _{\mu \nu }dx^{\mu }dx^{\nu }\rightarrow d\tau ^{\prime
2}=-\eta _{\mu \nu }\alpha _{\kappa }^{\mu }\alpha _{\lambda }^{\nu
}dx^{\kappa }dx^{\lambda }=-\eta _{\kappa \lambda }dx^{\kappa }dx^{\lambda
}=d\tau ^{2}.
\end{equation}%
In contrast, the Einstein group entails 16 instead of 10 independent
spacetime dependent parametric functions since in general 
\begin{equation}
g_{\mu \nu }x^{\mu },_{\kappa ^{\prime }}x_{,\lambda ^{\prime }}^{\nu
}=g_{\kappa ^{\prime }\lambda ^{\prime }}\neq g_{\kappa \lambda },
\end{equation}%
and therefore, unlike Eq. (\ref{eta}), does not restrict the $x_{,\kappa
^{\prime }}^{\mu }$ \ (in case of Lorentz transformations one has $%
x_{,\kappa ^{\prime }}^{\mu }=\alpha _{\kappa ^{\prime }}^{\mu })$ and of
course

\begin{equation}
d\tau ^{2}=-g^{\mu \nu }(x)dx_{\mu }dx_{\nu }=-g_{\mu \nu }(x)x^{\mu
},_{\kappa ^{\prime }}x_{,\lambda ^{\prime }}^{\nu }dx^{\prime \lambda
}dx^{\prime \kappa }=-g_{\kappa \lambda }^{\prime }(x)dx^{\prime \lambda
}dx^{\prime \kappa }=d\tau ^{\prime 2}.
\end{equation}%
In analogy with Dirac's idea of taking the square root of the Klein-Gordon
equation by introducing matrices, Sachs came upon the idea of taking the
square root of the metric and, in a sense, ultimately of the Einstein
equations themselves by using quaternions\cite{s1}-\cite{s3}. \ He does this
by introducing%
\begin{equation}
d\mathcal{S=}q_{\mu }(x)dx^{\mu },  \label{ds}
\end{equation}%
as a matrix square root of the squared line element instead of $\pm \sqrt{%
-d\tau ^{2}}$ $.~$

The quaternionic function $q^{\mu }(x)$ has both a vector character and a
second rank spinor character. \ That is, one can view it in terms of its
transformation properties as the outer product of two two-component spinors $%
\ \sim (\eta \eta ^{\dag })^{\mu }$ and so it transforms as a combination of
a four vector (first rank tensor) and as a second rank spinor under the
Einstein group, \ 
\begin{equation}
q_{\lambda ^{\prime }}^{\prime }(x^{\prime })=x^{\nu },_{\lambda ^{\prime
}}S(x)q_{\nu }S^{-1}(x),
\end{equation}%
where $S(x)$ are the spinor transformation matrices for the Einstein group.

\section{Square root of the metric condition.}

\ In addition to Eq. (\ref{ds}) Sachs introduces the time conjugate line
element 
\begin{equation}
d\mathcal{\tilde{S}}\mathcal{=}\tilde{q}_{\mu }(x)dx^{\mu },
\end{equation}%
where, as in Eq. (\ref{cj}), the quaternionic conjugate is defined by%
\begin{equation}
\tilde{q}_{\mu }(x)=\varepsilon q_{\mu }^{\ast }(x)\varepsilon .  \label{qc}
\end{equation}%
Then%
\begin{equation}
d\tau ^{2}=-d\mathcal{S}d\mathcal{\tilde{S}}=-q_{\mu }(x)\tilde{q}_{\nu
}(x)dx^{\mu }dx^{\nu }=-\sigma _{0}g_{\mu \nu }(x)dx^{\mu }dx^{\nu },
\end{equation}%
implies\cite{s4}, because of the symmetry in the differential indices,%
\begin{equation}
g_{\mu \nu }(x)\sigma _{0}=\frac{1}{2}[q_{\mu }(x)\tilde{q}_{\nu }(x)+q_{\nu
}(x)\tilde{q}_{\mu }(x)].  \label{qm}
\end{equation}%
For Minkowski space, or in the local \ limit, ~$g_{\mu \nu }\rightarrow \eta
_{\mu \nu }$ and $q_{\mu }(x)\rightarrow \sigma _{\mu }$, as one can readily
verify,%
\begin{equation}
\eta _{\mu \nu }\sigma _{0}=\frac{1}{2}[\sigma _{\mu }\tilde{\sigma}_{\nu
}+\sigma _{\nu }\tilde{\sigma}_{\mu }].
\end{equation}

One can see how the 16 independent quaternion components can be explicitly
labeled by introducing the 16 real tetrads $v_{a}^{\mu }(x)$ $\ $with one
Greek index in general space time and one Latin index in Minkowski space,%
\begin{equation}
q^{\mu }(x)=\eta ^{ab}\sigma _{b}v_{a}^{\mu }(x)=q^{\mu }(x)^{\dag },
\label{1}
\end{equation}%
with 
\begin{equation}
v_{a}^{\mu }(x)v_{b}^{\nu }(x)g_{\mu \nu }(x)=\eta _{ab},
\end{equation}%
and%
\begin{equation}
\eta ^{ab}v_{a}^{\mu }(x)v_{b}^{\nu }(x)=g^{\mu \nu }.
\end{equation}%
As with Eq. (\ref{qm}), this can be viewed as giving 10 conditions on the 16
functions embodied in the 4 tetrads. The conjugate quaternion is given by%
\begin{equation}
\tilde{q}^{\mu }(x)=\varepsilon q^{\mu }(x)^{\ast }\varepsilon =\eta ^{ab}%
\tilde{\sigma}_{b}v_{a}^{\mu }(x)=\tilde{q}^{\mu }(x)^{\dag }.  \label{2}
\end{equation}%
The Minkowski limit is defined by%
\begin{eqnarray}
v_{a}^{\mu }(x) &\rightarrow &\delta _{a}^{\mu },  \notag \\
v_{a\mu }(x) &\rightarrow &\eta _{a\mu }.
\end{eqnarray}%
Using Eq. (\ref{1}), (\ref{2}), and%
\begin{equation}
Tr\sigma _{a}\tilde{\sigma}_{b}=2\eta _{ab}
\end{equation}%
one can show that%
\begin{equation}
Trq^{\mu }(x)\tilde{q}^{\nu }(x)=2g^{\mu \nu }(x).
\end{equation}

\section{The Spin Affine Connection}

\ \ Just as a vector field's covariant derivative requires the introduction
of the affine connection $\Gamma _{\mu \nu }^{\kappa }$%
\begin{equation}
\Gamma _{\mu \nu }^{\kappa }=g^{\kappa \sigma }\Gamma _{\sigma \mu \nu }=%
\frac{g^{\kappa \sigma }}{2}(g_{\nu \sigma ,\mu }+g_{\mu \sigma ,\nu
}-g_{\mu \nu ,\sigma })=\Gamma _{\nu \mu }^{\kappa },
\end{equation}%
so a spinor field requires the introduction of the spin-affine connection%
\cite{berg}. \ Both affine connections are due to the nonlinear space-time.
\ It is important to note that the introduction of the 2x2 matrix structure
of the quaternion $q_{\mu }$ implies a spinor vector space upon which it can
act, with two component spinors as elements, 
\begin{equation}
\eta =%
\begin{bmatrix}
\eta _{1} \\ 
\eta _{2}%
\end{bmatrix}%
.
\end{equation}

Just as a four vector $V^{\mu }(x)$ under the Einstein group transforms as a
first rank vector, 
\begin{equation}
V^{\mu }(x)\rightarrow V^{\mu ^{\prime }}(x^{\prime })=x^{\mu ^{\prime
}},_{\nu }(x)V^{\mu }(x),
\end{equation}%
so the spinor $\eta (x)$ transforms as a first rank spinor, \ 
\begin{equation}
\eta (x)\rightarrow \eta ^{\prime }(x^{\prime })=S(x)\eta (x).
\end{equation}%
\ One thus anticipates a covariant derivative of the form%
\begin{equation}
\eta _{;\mu }=\eta ,_{\mu }+\Omega _{\mu }\eta .
\end{equation}%
This becomes more clear by making explicit the spinor index%
\begin{equation}
\eta _{;\mu }^{\alpha }=\eta ,_{\mu }^{\alpha }+\Omega _{\beta \mu }^{\alpha
}\eta ^{\beta },
\end{equation}%
in analogy with the way in which the ordinary affine connection modifies the
gradient of a vector to produce a covariant derivative, 
\begin{equation}
V_{;\mu }^{\nu }=V_{,\mu }^{\nu }+\Gamma _{\mu \lambda }^{\nu }V^{\kappa }.
\end{equation}%
\ 

One finds $\Omega _{\mu }$ \cite{berg} by noting that just as the metric has
a zero covariant derivative, $g_{\mu \nu ;\lambda }=0,$ so $q_{;\lambda
}^{\mu }=0=\tilde{q}_{;\lambda }^{\mu }$ because of the connection (\ref{qm}%
) between $q$ and $g$ or that in the local limit $q_{;\lambda }^{\mu
}\rightarrow \sigma _{;\lambda }^{\mu }=0$. \ That leaves open the question
of how to define the covariant derivative of an object that is at the same
time a vector and a second rank spinor.

One expects that, in analogy to the expression for the covariant derivative
of an ordinary third rank tensor,%
\begin{equation}
T_{~\ \ \ ;\lambda }^{\mu \nu \kappa }=T_{~\ \ \ ,\lambda }^{\mu \nu \kappa
}+\Gamma _{\lambda \eta }^{\mu }T^{\eta \nu \kappa }+\Gamma _{\lambda \eta
}^{\nu }T^{\mu \eta \kappa }+\Gamma _{\lambda \eta }^{\kappa }T^{\mu \nu
\eta },
\end{equation}%
that 
\begin{equation}
q_{;\lambda }^{\mu \alpha \beta }=q_{,\lambda }^{\mu \alpha \beta }+\Gamma
_{\tau \lambda }^{\mu }q^{\tau \alpha \beta }+\Omega _{\gamma \lambda
}^{\alpha }q^{\mu \gamma \beta }+q^{\mu \alpha \gamma }\Omega _{\gamma
\lambda }^{\ast \beta },
\end{equation}%
or\cite{berg}%
\begin{equation}
q_{;\lambda }^{\mu }=q^{\mu },_{\lambda }+\Omega _{\lambda }q^{\mu }+q^{\mu
}\Omega _{\lambda }^{\dag }+\Gamma _{\tau \lambda }^{\mu }q^{\tau }=0.
\end{equation}%
Contract with $\tilde{q}^{\mu },$%
\begin{equation}
0=\tilde{q}_{\mu }(q^{\mu },_{\lambda }+\Omega _{\lambda }q^{\mu }+q^{\mu
}\Omega _{\lambda }^{\dag }+\Gamma _{\tau \lambda }^{\mu }q^{\tau }),
\label{qd}
\end{equation}%
and use%
\begin{equation}
\tilde{q}_{\mu }Aq^{\mu }=4A^{0}\sigma _{0}=2TrA,  \label{qb}
\end{equation}%
and%
\begin{equation}
A\varepsilon +\varepsilon A^{T}=2\varepsilon TrA.  \label{qq}
\end{equation}%
Hence%
\begin{equation}
\Omega _{\lambda }\varepsilon +\varepsilon \Omega _{\lambda
}^{T}=\varepsilon Tr\Omega _{\lambda }.  \label{o}
\end{equation}%
Now since $\varepsilon $ is a second rank spinor tensor like $q$ we have%
\begin{equation}
\varepsilon _{;\lambda }=0=\varepsilon ,_{\lambda }+\Omega _{\lambda
}\varepsilon +\varepsilon \Omega _{\lambda }^{T}=\Omega _{\lambda
}\varepsilon +\varepsilon \Omega _{\lambda }^{T}.
\end{equation}%
\ Thus, \ from Eq. (\ref{qq})%
\begin{equation}
Tr\Omega _{\lambda }=0,
\end{equation}%
and from Eq. (\ref{qb}) 
\begin{equation}
\tilde{q}_{\mu }\Omega _{\lambda }q^{\mu }=0,
\end{equation}%
so that\cite{berg}%
\begin{eqnarray}
\Omega _{\lambda }^{\dag } &=&-\frac{1}{4}\tilde{q}_{\mu }(q^{\mu
},_{\lambda }+\Gamma _{\tau \lambda }^{\mu }q^{\tau }),  \notag \\
\Omega _{\lambda } &=&\varepsilon \left( \Omega _{\lambda }^{\dag }\right)
^{\ast }\varepsilon =\frac{1}{4}\varepsilon \tilde{q}_{\mu }^{\ast
}\varepsilon \varepsilon (q^{\mu },_{\lambda }\varepsilon +\Gamma _{\tau
\lambda }^{\mu }q^{\tau }\varepsilon )=\frac{1}{4}q_{\mu }(\tilde{q}^{\mu
},_{\lambda }+\Gamma _{\tau \lambda }^{\mu }\tilde{q}^{\tau }),  \label{52}
\end{eqnarray}%
Taking the adjoint, gives us the two additional forms 
\begin{eqnarray}
\Omega _{\lambda }^{\dag } &=&\frac{1}{4}(\tilde{q}^{\mu },_{\lambda
}+\Gamma _{\tau \lambda }^{\mu }\tilde{q}^{\tau })q_{\mu },  \notag \\
\Omega _{\lambda } &=&-\frac{1}{4}(q^{\mu },_{\lambda }+\Gamma _{\tau
\lambda }^{\mu }q^{\tau })\tilde{q}_{\mu }.  \label{o3}
\end{eqnarray}

\section{The Riemann Curvature Tensor, the Spin Curvature Tensor and Their
Relation}

For an arbitrary first rank tensor $A_{\nu }$ the mixed second covariant
derivatives do not commute, with their difference 
\begin{eqnarray*}
A_{\nu ;\rho ;\sigma }-A_{\nu ;\sigma ;\rho } &=&(\Gamma _{\nu \sigma
}^{\kappa },_{\rho }-\Gamma _{\nu \rho }^{\kappa },_{\sigma })A_{\kappa
}+(\Gamma _{\nu \sigma }^{\kappa }\Gamma _{\kappa \rho }^{\lambda }-\Gamma
_{\nu \rho }^{\kappa }\Gamma _{\kappa \sigma }^{\lambda })A_{\lambda } \\
&=&R_{\nu \sigma \rho }^{\lambda }A_{\lambda }=R_{\lambda \nu \sigma \rho
}A^{\lambda },
\end{eqnarray*}%
defining the fourth rank mixed Riemann Christoffel curvature tensor. 
\begin{equation}
R_{\nu \rho \sigma }^{\lambda }=\Gamma _{\nu \sigma ,\rho }^{\lambda
}-\Gamma _{\nu \rho ,\sigma }^{\lambda }+\Gamma _{\nu \sigma }^{\kappa
}\Gamma _{\kappa \rho }^{\lambda }-\Gamma _{\nu \rho }^{\kappa }\Gamma
_{\kappa \sigma }^{\lambda },  \label{113}
\end{equation}

\bigskip In analogy to this use the fact that $\eta _{;\rho }=\eta ,_{\rho
}+\Omega _{\rho }\eta $ is both a first rank tensor and a first rank spinor$%
~ $from which one obtains%
\begin{eqnarray}
\eta _{;\rho ;\lambda } &=&\left( \eta ,_{\rho }+\Omega _{\rho }\eta \right)
,_{\lambda }-\Gamma _{\rho \lambda }^{\nu }\left( \eta ,_{\nu }+\Omega _{\nu
}\eta \right) +\Omega _{\lambda }\left( \eta ,_{\rho }+\Omega _{\rho }\eta
\right)  \notag \\
&=&\eta ,_{\rho },_{\lambda }+\Omega _{\rho ,\lambda }\eta +\Omega _{\rho
}\eta ,_{\lambda }-\Gamma _{\rho \lambda }^{\nu }\left( \eta ,_{\nu }+\Omega
_{\nu }\eta \right) +\Omega _{\lambda }\left( \eta ,_{\rho }+\Omega _{\rho
}\eta \right) ,
\end{eqnarray}%
and so\cite{berg} 
\begin{eqnarray}
\eta _{;\rho ;\lambda }-\eta _{;\lambda ;\rho } &=&[\Omega _{\rho
},_{\lambda }+\Omega _{\lambda }\Omega _{\rho }-\Omega _{\lambda },_{\rho
}-\Omega _{\rho }\Omega _{\lambda }]\eta  \notag \\
&\equiv &K_{\lambda \rho }\eta ,  \notag \\
K_{\lambda \rho } &=&\Omega _{\rho },_{\lambda }+\Omega _{\lambda }\Omega
_{\rho }-\Omega _{\lambda },_{\rho }-\Omega _{\rho }\Omega _{\lambda }.
\label{om}
\end{eqnarray}%
$K_{\lambda \rho }$ denotes the spin curvature tensor. Similarly 
\begin{eqnarray}
\eta _{;\rho ;\lambda }^{\dag }-\eta _{;\lambda ;\rho }^{\dag } &=&\eta
^{\dag }K_{\lambda \rho }^{\dag },  \notag \\
K_{\lambda \rho }^{\dag } &=&\Omega _{\rho ,\lambda }^{\dag }+\Omega _{\rho
}^{\dag }\Omega _{\lambda }^{\dag }-\Omega _{\lambda ,\rho }^{\dag }-\Omega
_{\lambda }^{\dag }\Omega _{\rho }^{\dag }.  \label{mo}
\end{eqnarray}%
Note that from Eq. (\ref{52}) 
\begin{equation}
\varepsilon K_{\lambda \rho }^{\ast }\varepsilon =\varepsilon (\Omega _{\rho
,\lambda }\varepsilon -\Omega _{\lambda }^{\ast }\varepsilon \varepsilon
\Omega _{\rho }^{\ast }\varepsilon -\varepsilon \Omega _{\lambda ,\rho
}^{\ast }\varepsilon +\varepsilon \Omega _{\rho }^{\ast }\varepsilon
\varepsilon \Omega _{\lambda }^{\ast }\varepsilon )=K_{\lambda \rho }^{\dag }
\label{K}
\end{equation}

Appendix A demonstrates, by using the above two connections between mixed
covariant derivatives and the spin curvature tensor, that\cite{berg}%
\begin{eqnarray}
K_{\rho \lambda }q_{\mu }+q_{\mu }K_{\rho \lambda }^{\dag } &=&-R_{\kappa
\mu \rho \lambda }q^{\kappa },  \notag \\
K_{\rho \lambda }^{\dag }\tilde{q}_{\mu }+\tilde{q}_{\mu }K_{\rho \lambda }
&=&R_{\kappa \mu \rho \lambda }\tilde{q}^{\kappa }.  \label{rk}
\end{eqnarray}%
Given the forms in Eq. (\ref{om}), (\ref{mo}) \ and (\ref{52}-\ref{o3}) on
the one hand and Eq. (\ref{113}) on the other, this equation is plausible
because of the connection between the metric and the quaternions given in
Eq. (\ref{qm}) and the fact that the curvature tensor on the right hand side
involves first and second derivatives of the metric tensor through the
affine connection while the spin curvature tensor on the left hand side
involves first and second derivative of the quaternions $q_{\mu }$ and $%
\tilde{q}_{\mu }$ through the spin affine connection. As far as we have been
able to determine, however, there has been no published proof of Eq. (\ref%
{rk}) by manipulations involving traces say of the sort\footnote{%
The missing technology appears to be the analogue of the traces involving 4
and 6 \ gamma matrices when derivatives are involved. \ One could mimic the
gamma matrix proofs by use of the tetrad relations in Eq. (\ref{1}) and Eq. (%
\ref{2}), but the expressions involving the derivatives of the $q^{\prime }s$
in the expression for $K$ complicates attempts to show explicitly that
evaluation of the left hand side of Eq. (\ref{no}) yields the right hand
side.} 
\begin{equation}
-\frac{1}{2}Tr[(K_{\rho \lambda }q_{\mu }+q_{\mu }K_{\rho \lambda }^{\dag })%
\tilde{q}_{\eta }]=\frac{1}{2}R_{\kappa \mu \rho \lambda }Trq^{\kappa }%
\tilde{q}_{\eta }=R_{\eta \mu \rho \lambda }.  \label{no}
\end{equation}%
\ \ 

Let us multiply the first of Eqs. (\ref{rk}) by $\tilde{q}^{\mu }$ on the
right \ and the second by $q^{\mu }$ on the left and add the two
expressions. \ One obtains%
\begin{equation}
K_{\rho \lambda }q_{\mu }\tilde{q}^{\mu }+q_{\mu }K_{\rho \lambda }^{\dag }%
\tilde{q}^{\mu }+q^{\mu }K_{\rho \lambda }^{\dag }\tilde{q}_{\mu }+q^{\mu }%
\tilde{q}_{\mu }K_{\rho \lambda }=R_{\kappa \mu \rho \lambda }(q^{\mu }%
\tilde{q}^{\kappa }-q^{\kappa }\tilde{q}^{\mu }).
\end{equation}%
To simplify this one uses $q^{\mu }A\tilde{q}_{\mu }=2TrA,$ $q^{\mu }\tilde{q%
}_{\mu }=2Tr1=4$, $Tr\Omega _{\lambda }=0\ $and%
\begin{eqnarray}
TrK_{\lambda \rho } &=&Tr[\partial _{\lambda }\Omega _{\rho }+\Omega
_{\lambda }\Omega _{\rho }-\partial _{\rho }\Omega _{\lambda }-\Omega _{\rho
}\Omega _{\lambda }]  \notag \\
&=&Tr[\partial _{\lambda }\Omega _{\rho }-\partial _{\rho }\Omega _{\lambda
}]=\partial _{\lambda }Tr\Omega _{\rho }-\partial _{\rho }Tr\Omega _{\lambda
}=0.  \label{trk}
\end{eqnarray}%
Hence one finds that the following simple connection between the spin and
Riemann curvature tensor,\cite{berg}%
\begin{eqnarray}
K_{\rho \lambda } &=&\frac{1}{4}R_{\lambda \rho \mu \kappa }q^{\mu }\tilde{q}%
^{\kappa },  \notag \\
K_{\rho \lambda }^{\dag } &=&\frac{1}{4}R_{\lambda \rho \mu \kappa }\tilde{q}%
^{\kappa }q^{\mu }.  \label{kr}
\end{eqnarray}%
Even though these two equations as the two in (\ref{rk}) demonstrate
formally the connection between the spin and Riemann curvature tensors there
has been no verification of their equivalence by using just the definitions
of $K$ in terms of $\Omega $ and ultimately $q$ and its derivatives. \ We
will not attempt this here. \ Instead, we shall concern ourselves with
finding a solution to the quaternionic field equations.

\section{Sachs' Quaternionic Field Equation}

The Einstein equation is written in terms of the symmetric Ricci tensor%
\begin{equation}
R_{\nu \kappa }=R_{\kappa \nu }=R_{\nu \kappa \lambda }^{\mu }g_{\mu
}^{\lambda }=R_{\nu \kappa \mu }^{\mu },
\end{equation}%
\ and the scalar curvature,%
\begin{equation}
R=g^{\nu \kappa }R_{\nu \kappa }=R_{\kappa }^{\kappa }.
\end{equation}%
In the presence of matter or electromagnetic fields, it is%
\begin{equation}
G^{\mu \nu }=R^{\mu \nu }-\frac{1}{2}g^{\mu \nu }R=-8\pi T^{\mu \nu }.
\label{eine}
\end{equation}%
The source term $T^{\mu \nu },$ corresponding to the density and flux of
nongravitational energy and momentum, must satisfy 
\begin{equation}
T_{~~;\mu }^{\mu \nu }=0,
\end{equation}%
as does 
\begin{equation}
G_{~~;\mu }^{\mu \nu }=0
\end{equation}%
\ corresponding to the contracted Bianci identity.

As is well known\cite{dirac}, the Einstein equation also follows from the
action principle applied to%
\begin{equation}
I=\int \mathcal{L}_{E}d^{4}x,
\end{equation}%
in which the Lagrange function is%
\begin{equation}
\mathcal{L}_{E}\mathcal{=}g_{\mu \nu }R^{\mu \nu }\sqrt{-g}  \label{ea}
\end{equation}%
where $g$ is the determinant of the metric tensor. \ Including matter terms $%
\mathcal{L}_{M}$ and applying the action principle to $\mathcal{L}_{E}+%
\mathcal{L}_{M}$ gives Eq. (\ref{eine}).Sachs derives his quaternionic field
equation from a similar action, but with a quaternionic version of Eq. (\ref%
{ea}).

To find the quaternionic \ version of $R$ he shows first (see Appendix A)
that%
\begin{eqnarray}
\sigma _{0}R_{\gamma \mu \rho \lambda } &=&-\frac{1}{2}[\tilde{q}_{\gamma
}K_{\rho \lambda }q_{\mu }+\tilde{q}_{\gamma }q_{\mu }K_{\rho \lambda
}^{\dag }-K_{\rho \lambda }^{\dag }\tilde{q}_{\mu }q_{\gamma }-\tilde{q}%
_{\mu }K_{\rho \lambda }q_{\gamma }]\equiv \mathcal{R}_{\gamma \mu \rho
\lambda }  \notag \\
\sigma _{0}R_{\mu \rho } &=&\sigma _{0}g^{\gamma \lambda }R_{\gamma \mu \rho
\lambda }=-\frac{1}{2}[\tilde{q}^{\lambda }K_{\rho \lambda }q_{\mu }+\tilde{q%
}^{\lambda }q_{\mu }K_{\rho \lambda }^{\dag }-K_{\rho \lambda }^{\dag }%
\tilde{q}_{\mu }q^{\lambda }-\tilde{q}_{\mu }K_{\rho \lambda }q^{\lambda
}]\equiv \mathcal{R}_{\mu \rho },  \notag \\
\sigma _{0}R &=&g^{\mu \rho }\sigma _{0}R_{\mu \rho }=-\frac{1}{2}[\tilde{q}%
^{\lambda }K_{\rho \lambda }q^{\rho }+\tilde{q}^{\lambda }q^{\rho }K_{\rho
\lambda }^{\dag }-K_{\rho \lambda }^{\dag }\tilde{q}^{\rho }q^{\lambda }-%
\tilde{q}^{\rho }K_{\rho \lambda }q^{\lambda }]\equiv \mathcal{R}.
\label{sacre}
\end{eqnarray}%
We stress the difference between the number field forms on the left hand
side and the quaternion field forms on the right hand side by the use of
Roman and script variables. \ Taking the trace of the scalar curvature,%
\begin{equation*}
R=\frac{1}{2}Tr\mathcal{R}=\frac{1}{4}Tr[\tilde{q}^{\lambda }K_{\rho \lambda
}q^{\rho }+h.c.].
\end{equation*}%
For the Lagrangian density Sachs thus uses, in analogy to Eq. (\ref{ea}),%
\begin{equation}
\mathcal{L}_{E}=(Tr\mathcal{R})(-g)^{1/2}=\frac{1}{2}Tr[\tilde{q}^{\lambda
}K_{\rho \lambda }q^{\rho }+h.c.](-g)^{1/2}.
\end{equation}%
Sachs derives his quaternionic form of the metrical field equations in which 
$\mathcal{R}$ is regarded as a function of $q^{\mu }$ and $\tilde{q}^{\mu }$
by way of the right hand side from

\begin{eqnarray}
&&\delta \int \{Tr[\mathcal{R}(q^{\mu },\tilde{q}^{\mu },\Omega _{\mu
},\Omega _{\mu }^{\dag })]+\mathcal{L}_{M}\}(-g)^{1/2}d^{4}x,  \notag \\
\mathcal{L}_{M} &=&\text{(matter and electromagnetic contributions).}
\end{eqnarray}%
As with the method devised by Palatini in which the form of the affine
connection is not assumed but instead an outcome of the equations of motion,
so the spin affine connection $\Omega _{\mu }$ and its relation to the
derivatives of $q$ is an outcome of the equations of motion. \ This is
accomplished by regarding $\Omega $ as an independent variable. Using the
Palatini-like method, in which $\mathcal{L}_{E}$ depends on the spin
curvature $K_{\mu \nu }$ only through $\Omega _{\mu ;\nu }$ one finds%
\begin{eqnarray}
\delta \int \mathcal{L}_{E}d^{4}x &=&\int \left[ \frac{\partial \mathcal{L}%
_{E}}{\partial K_{\mu \nu }}\left( \frac{\partial K_{\mu \nu }}{\partial
\Omega _{\mu ;\nu }}\right) \right] _{;\mu }\delta \Omega _{\mu }d^{4}x=0, 
\notag \\
&\Longrightarrow &\left( \frac{\partial \mathcal{L}_{E}}{\partial K_{\mu \nu
}}\right) _{;\mu }=(q^{\nu }\tilde{q}^{\mu }-q^{\mu }\tilde{q}^{\nu })_{;\mu
}=0,
\end{eqnarray}%
which in turn leads to the relation derived above%
\begin{equation}
\Omega _{\lambda }=-\frac{1}{4}(q^{\mu },_{\lambda }+\Gamma _{\tau \lambda
}^{\mu }q^{\tau })\tilde{q}_{\mu },
\end{equation}%
between $\Omega _{\mu }$ and the quaternion $q^{\mu }$, its derivatives and
the affine connection. \ Having established this, using the general
relativistic Lagrange equations of motion,%
\begin{equation}
\frac{\partial \mathcal{L}}{\partial \Lambda ^{(i)}}=\left[ \frac{\partial 
\mathcal{L}}{\partial \Lambda _{;\mu }^{(i)}}\right] _{;\mu },
\end{equation}%
and $K_{\rho \lambda }=-$ $K_{\lambda \rho }$ leads to \cite{s6} 
\begin{eqnarray}
\frac{\partial \mathcal{L}_{E}}{\partial q^{\rho }} &=&\frac{1}{4}[-(K_{\rho
\lambda }^{\dag }\tilde{q}^{\lambda }+\tilde{q}^{\lambda }K_{\rho \lambda })-%
\frac{1}{2}\mathcal{R}\tilde{q}_{\rho }]^{\ast }(-g)^{1/2},  \notag \\
\frac{\partial \mathcal{L}_{E}}{\partial \tilde{q}^{\rho }} &=&\frac{1}{4}%
[(K_{\rho \lambda }q^{\lambda }+q^{\lambda }K_{\rho \lambda }^{\dag })-\frac{%
1}{2}\mathcal{R}q_{\rho }]^{\ast }(-g)^{1/2}
\end{eqnarray}%
and so 
\begin{eqnarray}
\frac{1}{4}(K_{\rho \lambda }q^{\lambda }+q^{\lambda }K_{\rho \lambda
}^{\dag })-\frac{1}{8}\mathcal{R}q_{\rho } &=&-\frac{\partial \mathcal{L}_{M}%
}{\partial \tilde{q}^{\rho }}=k\mathcal{F}_{\rho },  \notag \\
-\frac{1}{4}(K_{\rho \lambda }^{\dag }\tilde{q}^{\lambda }+\tilde{q}%
^{\lambda }K_{\rho \lambda })-\frac{1}{8}\mathcal{R}\tilde{q}_{\rho } &=&k%
\mathcal{\tilde{F}}_{\rho }=k\varepsilon \mathcal{F}_{\rho }^{\ast
}\varepsilon .  \label{fac}
\end{eqnarray}%
Using Eq. (\ref{K}), one sees that these two equations are quaternionic
conjugates of one another. Each of these equations transform as a vector
quaternion with 16 independent components. Either of these nonlinear second
order partial differential equations is sufficient to determine fully the 16
independent parts of the quaternion $q^{\mu }(x)$ given appropriate boundary
conditions$.$ It is appropriate to call this equation and its conjugate the
Sachs equations. But before we go on to construct a quaternionic solution to
these Sachs equations let us make some remarks about their intrinsic lack of
either time reversal symmetry or space inversion. \ 

Recall that quaternionic conjugation Eq. (\ref{qc}) is equivalent to time
reversal. \ The two equations in (\ref{fac}) are therefore temporal
reflections of each other but are distinct and independent. One could
likewise show that in Eq. (\ref{fac}) the two equations are spatial
reflections of one another.\ The separation into conjugated field equations
appears because of the lack of reflection symmetry in the Einstein group.
These two equations, from a mathematical point of view, are analogous to two
complex equations, say $f(z^{\ast },z)=0$ and $f(z,z^{\ast })=0,$ which are
complex conjugates of one another assuming $f$ is a real function. These are
similar to what Sachs has with the two equations in (\ref{fac}) which are
the quaternionic conjugates of each other. Since quaternionic conjugation is
the same as either spatial or temporal inversion, his equations go into each
other under either parity or time reversal. One could obtain an even parity
equation from Eq. (\ref{fac}) by adding the two equations, in analogy to
using the real equation $f(z^{\ast },z)+f(z,z^{\ast })$ instead of the
separate complex equations. \ Likewise one could obtain an odd parity
equation from Eq. (\ref{fac}) by subtracting the two equations, in analogy
to using the imaginary equation $f(z^{\ast },z)-f(z,z^{\ast })$ instead of
the separate complex equations. Note this absence of reflection symmetry of
each of the two equations in (\ref{fac}) is, however, not the same as parity
or time reversal violation in a \textit{single} equation because the physics
involves not just one of the equations, but both. \ 

\section{\ An Exact Solution to the Vacuum Sachs Equation.}

The most well known solution of the Einstein equation is the Schwarzschild
solution. \ It is an exact solution. \ In this section we will demonstrate a
similar exact solution to the Sachs equation (\ref{fac}). \ But first, we
review the standard form of the Schwarzschild solution of the Einstein
equation in a vacuum with spherical symmetry and static conditions . \ Let 
\begin{eqnarray}
x^{0} &=&t,  \notag \\
x^{1} &=&r,  \notag \\
x^{2} &=&\theta ,  \notag \\
x^{3} &=&\phi .
\end{eqnarray}%
Using Dirac's form of the metric,\cite{dirac},%
\begin{eqnarray}
d\tau ^{2} &=&e^{2\nu }dt^{2}-e^{2\lambda }dr^{2}-r^{2}(d\theta ^{2}+\sin
^{2}\theta d\phi ^{2}),  \notag \\
g_{00} &=&-e^{2\nu }=1/g^{00}  \notag \\
g_{11} &=&e^{2\lambda }=1/g^{11},  \notag \\
g_{22} &=&r^{2}=1/g^{22},  \notag \\
g_{33} &=&r^{2}\sin ^{2}\theta =1/g^{33}.  \notag \\
g_{\mu \nu } &=&0,~\mu \neq \nu .
\end{eqnarray}%
Recall that 
\begin{equation}
\Gamma _{\mu \nu }^{\kappa }=\frac{g^{\kappa \sigma }}{2}(g_{\nu \sigma ,\mu
}+g_{\mu \sigma ,\nu }-g_{\mu \nu ,\sigma })=\Gamma _{\nu \mu }^{\kappa }.
\end{equation}%
The only nonzero $\Gamma ^{\prime }s$ are \cite{dirac}%
\begin{eqnarray}
\Gamma _{00}^{1} &=&\nu ^{\prime }e^{2\nu -2\lambda },~~~~~~~~\Gamma
_{10}^{0}=\nu ^{\prime },  \notag \\
\Gamma _{11}^{1} &=&\lambda ^{\prime },~~~~~~~~~~~~~~~~~\Gamma
_{12}^{2}=\Gamma _{13}^{3}=r^{-1},  \notag \\
\Gamma _{22}^{1} &=&-re^{-2\lambda },~~~~~~~~~~\Gamma _{23}^{3}=\cot \theta ,
\notag \\
\Gamma _{33}^{1} &=&-r\sin ^{2}\theta e^{-2\lambda },~~\Gamma
_{33}^{2}=-\sin \theta \cos \theta .  \label{afin}
\end{eqnarray}%
With%
\begin{equation}
R_{\nu \sigma }=\Gamma _{\nu \lambda }^{\lambda },_{\sigma }-\Gamma _{\nu
\sigma }^{\lambda },_{\lambda }+\Gamma _{\nu \sigma }^{\kappa }\Gamma
_{\kappa \lambda }^{\lambda }-\Gamma _{\nu \kappa }^{\kappa }\Gamma _{\sigma
\lambda }^{\lambda },  \label{ric}
\end{equation}%
the vacuum Einstein equation $R_{\nu \sigma }=0$ are the Schwarzschild
equations:%
\begin{eqnarray}
R_{00} &=&\left( -\nu ^{\prime \prime }+\lambda ^{\prime }\nu ^{\prime }-\nu
^{\prime 2}-\frac{2\nu ^{\prime }}{r}\right) e^{2\nu -2\lambda }=0,  \notag
\\
R_{11} &=&\nu ^{\prime \prime }-\lambda ^{\prime }\nu ^{\prime }+\nu
^{\prime 2}-\frac{2\lambda ^{\prime }}{r}=0,  \notag \\
R_{22} &=&(1+r\nu ^{\prime }-r\lambda ^{\prime })e^{-2\lambda }-1=0,  \notag
\\
R_{33} &=&R_{22}\sin ^{2}\theta =0.  \label{schw}
\end{eqnarray}%
These lead to%
\begin{eqnarray}
\lambda ^{\prime } &=&-\nu ^{\prime },  \notag \\
\nu ^{\prime \prime }+2\nu ^{\prime 2}+\frac{2\nu ^{\prime }}{r} &=&0, 
\notag \\
(1+2r\nu ^{\prime })e^{2\nu }-1 &=&0.  \label{3s}
\end{eqnarray}%
They are sufficient, together with matching at large $r$ onto the Newtonian
form to determine the solutions, 
\begin{eqnarray}
g_{00} &=&-1+\frac{r_{s}}{r}=1/g^{00}  \notag \\
g_{11} &=&e^{2\lambda }=1/(1-\frac{r_{s}}{r})=1/g^{11},  \notag \\
g_{22} &=&r^{2}=1/g^{22},  \notag \\
g_{33} &=&r^{2}\sin ^{2}\theta =1/g^{33}.
\end{eqnarray}%
in which $r_{s}=2GM$ is the Schwarzschild radius with $G$ the gravitational
constant and $M$ the gravitating mass (we use units with $c=1$) .

\bigskip We consider the quaternionic four-vector equation given in Eq. (\ref%
{fac}), the Sachs equation. \ The vacuum form of this equation is\footnote{%
Stricltly speaking we are solving the equation for $r\neq 0$ or outside the
source and matching the solution to Newton's for large $r$. \ Sachs points
out \cite{s3} that the Schwarzschild solution should be a valid
approximation for a bona fide nonlinear solution of the full non-homogeneous
Einstein equation, though this has yet to be demonstrated analytically. \ } 
\begin{equation}
\frac{1}{4}(K_{\rho \lambda }q^{\lambda }+q^{\lambda }K_{\rho \lambda
}^{\dag })-\frac{1}{8}\mathcal{R}q_{\rho }=0  \label{vacsac}
\end{equation}%
Now, we seek to find the quaternionic solution of the Sachs equation
corresponding to the above metric. \ Consider the ansatz%
\begin{eqnarray}
q_{0} &=&e^{\nu }~,~~~~~~~~~~~q^{0}=-e^{-\nu },  \notag \\
q_{1} &=&\mathbf{\hat{x}\cdot \sigma }e^{\lambda },~~~~~q^{1}=\mathbf{\hat{x}%
\cdot \sigma }e^{-\lambda },  \notag \\
q_{2} &=&r\mathbf{\hat{\theta}\cdot \sigma },~~~~~~~~q^{2}=\frac{\mathbf{%
\hat{\theta}\cdot \sigma }}{r},  \notag \\
q_{3} &=&\sin \theta r\mathbf{\hat{\phi}\cdot \sigma ;~~}q^{3}=\frac{\mathbf{%
\hat{\phi}\cdot \sigma }}{\sin \theta r}.  \label{qstz}
\end{eqnarray}%
They are plausible since they satisfy%
\begin{equation}
q^{\mu }\tilde{q}^{\nu }+q^{\nu }\tilde{q}^{\mu }=2g^{\mu \nu }\sigma _{0}.
\end{equation}%
However, this does not necessarily imply that these satisfy the Sachs
metrical field equation (\ref{fac}). Eq. (\ref{qstz}) parametrizes the
tetrads of Eq. (\ref{1}) by%
\begin{eqnarray}
v_{0}^{0}(x) &=&e^{-\nu },  \notag \\
v_{1}^{1}(x) &=&e^{-\lambda },  \notag \\
v_{2}^{2}(x) &=&\frac{1}{r},  \notag \\
v_{3}^{3}(x) &=&\frac{1}{r\sin \theta },  \notag \\
v_{a}^{\mu }(x) &=&0,~\mu \neq a.  \label{tetrad}
\end{eqnarray}

$\ \ \ $To compute the terms of the Sachs equation (\ref{fac}) consider first%
\begin{equation}
K_{\rho \lambda }=\Omega _{\rho },_{\lambda }+\Omega _{\lambda }\Omega
_{\rho }-\Omega _{\lambda },_{\rho }-\Omega _{\rho }\Omega _{\lambda },
\end{equation}%
where%
\begin{equation}
\Omega _{\lambda }=-\frac{1}{4}(q^{\mu },_{\lambda }+\Gamma _{\tau \lambda
}^{\mu }q^{\tau })\tilde{q}_{\mu }.
\end{equation}%
From this we have%
\begin{eqnarray}
\Omega _{\rho },_{\lambda } &=&-\frac{1}{4}\partial _{\lambda }\left[
(q^{\mu },_{\rho }+\Gamma _{\tau \rho }^{\mu }q^{\tau })\tilde{q}_{\mu }%
\right]  \notag \\
&=&-\frac{1}{4}[(q^{\mu },_{\rho },_{\lambda }+\Gamma _{\tau \rho ,\lambda
}^{\mu }q^{\tau }+\Gamma _{\tau \rho }^{\mu }q^{\tau },_{\tau })\tilde{q}%
_{\mu }  \notag \\
&&+(q^{\mu },_{\rho }+\Gamma _{\tau \rho }^{\mu }q^{\tau })\tilde{q}_{\mu
},_{\lambda }].
\end{eqnarray}%
Now, for the static case, there is no time dependence so that%
\begin{eqnarray}
K_{0l} &=&\Omega _{0,l}+\Omega _{l}\Omega _{0}-\Omega _{0}\Omega _{l}. 
\notag \\
\Omega _{0} &=&-\frac{1}{4}\Gamma _{\tau 0}^{\mu }q^{\tau }\tilde{q}_{\mu },
\notag \\
\Omega _{l} &=&-\frac{1}{4}(q^{\mu },_{l}+\Gamma _{\tau l}^{\mu }q^{\tau })%
\tilde{q}_{\mu },
\end{eqnarray}%
and using the only nonzero components of the affine connection from Eq. (\ref%
{afin}) and the ansatz Eq. (\ref{qstz}) we find 
\begin{eqnarray*}
\Omega _{0} &=&-\frac{1}{4}\Gamma _{\tau 0}^{\mu }q^{\tau }\tilde{q}_{\mu }=%
\frac{\nu ^{\prime }\mathbf{\hat{x}\cdot \sigma }e^{\nu -\lambda }}{2}, \\
\Omega _{1} &=&-\frac{1}{4}(q^{\mu },_{1}+\Gamma _{\tau 1}^{\mu }q^{\tau })%
\tilde{q}_{\mu }=0, \\
\Omega _{2} &=&-\frac{1}{4}(q^{\mu },_{2}+\Gamma _{\tau 2}^{\mu }q^{\tau })%
\tilde{q}_{\mu }=\frac{i}{2}(1-e^{-\lambda })\mathbf{\phi \cdot \sigma ,} \\
\Omega _{3} &=&-\frac{1}{4}(q^{\mu },_{3}+\Gamma _{\tau 3}^{\mu }q^{\tau })%
\tilde{q}_{\mu }=-\frac{i}{2}(1-e^{-\lambda })\sin \theta \mathbf{\theta
\cdot \sigma .}
\end{eqnarray*}%
Thus%
\begin{eqnarray}
K_{01} &=&\Omega _{0},_{1}+\Omega _{1}\Omega _{0}-\Omega _{0}\Omega _{1}=%
\frac{(\nu ^{\prime \prime }+\nu ^{\prime 2}-\nu ^{\prime }\lambda ^{\prime
})\mathbf{\hat{x}\cdot \sigma }e^{\nu -\lambda }}{2},  \notag \\
K_{02} &=&\Omega _{0},_{2}+\Omega _{2}\Omega _{0}-\Omega _{0}\Omega _{2}=%
\frac{\nu ^{\prime }\mathbf{\hat{\theta}\cdot \sigma }e^{\nu -2\lambda }}{2},
\notag \\
K_{03} &=&\Omega _{0},_{3}+\Omega _{3}\Omega _{0}-\Omega _{0}\Omega _{3}=%
\mathbf{\hat{\phi}\cdot \sigma }\frac{\nu ^{\prime }\sin \theta e^{\nu
-2\lambda }}{2},  \notag \\
K_{12} &=&\Omega _{1},_{2}+\Omega _{2}\Omega _{1}-\Omega _{2},_{1}-\Omega
_{1}\Omega _{2}\mathbf{=}-\frac{i\lambda ^{\prime }}{2}e^{-\lambda }\mathbf{%
\phi \cdot \sigma ,}  \notag \\
K_{13} &=&\Omega _{1},_{3}+\Omega _{3}\Omega _{1}-\Omega _{3},_{1}-\Omega
_{1}\Omega _{3}\mathbf{=}\frac{i}{2}\lambda ^{\prime }e^{-\lambda }\sin
\theta \mathbf{\theta \cdot \sigma ,}  \notag \\
K_{23} &=&\Omega _{2},_{3}-\Omega _{3},_{2}+\Omega _{3}\Omega _{2}-\Omega
_{2}\Omega _{3}=-\frac{i\sin \theta \mathbf{\hat{x}\cdot \sigma }}{2}%
(1-e^{-2\lambda }).
\end{eqnarray}

\bigskip

For our quaternionic ansatz (\ref{qstz}) we use these expressions for the
spin curvature tensors. \ Omitting details we obtain%
\begin{eqnarray}
K_{0\lambda }q^{\lambda } &=&K_{01}q^{1}+K_{02}q^{2}+K_{03}q^{3}  \notag \\
&=&\frac{e^{\nu -2\lambda }}{2}[\nu ^{\prime \prime }-\lambda ^{\prime }\nu
^{\prime }+\nu ^{\prime 2}+\frac{2\nu ^{\prime }}{r}]=\left( K_{0\lambda
}q^{\lambda }\right) ^{\dag }=q^{\lambda }K_{0\lambda }^{\dag },  \label{s0}
\end{eqnarray}%
and%
\begin{eqnarray}
K_{1\lambda }q^{\lambda } &=&K_{10}q^{0}+K_{12}q^{2}+K_{13}q^{3}  \notag \\
&=&\frac{e^{-\lambda }}{2}[\nu ^{\prime \prime }-\lambda ^{\prime }\nu
^{\prime }+\nu ^{\prime 2}-\frac{2\lambda ^{\prime }}{r}]\mathbf{\hat{x}%
\cdot \sigma }=\left( K_{1\lambda }q^{\lambda }\right) ^{\dag }=q^{\lambda
}K_{1\lambda }^{\dag },  \label{s1}
\end{eqnarray}%
and%
\begin{eqnarray}
K_{2\lambda }q^{\lambda } &=&K_{20}q^{0}+K_{21}q^{1}+K_{23}q^{3}  \notag \\
~ &=&\frac{e^{-2\lambda }}{2}[\nu ^{\prime }-\lambda ^{\prime }-\frac{%
(e^{2\lambda }-1)}{r}]\mathbf{\hat{\theta}\cdot \sigma }=\left( K_{2\lambda
}q^{\lambda }\right) ^{\dag }=q^{\lambda }K_{2\lambda }^{\dag },  \label{s2}
\end{eqnarray}%
and finally%
\begin{eqnarray}
K_{3\lambda }q^{\lambda } &=&K_{30}q^{0}+K_{31}q^{1}+K_{32}q^{2}  \notag \\
~ &=&\frac{\sin \theta e^{-2\lambda }}{2}[\nu ^{\prime }-\lambda ^{\prime }-%
\frac{(e^{2\lambda }-1)}{r}]\mathbf{\hat{\phi}\cdot \sigma }=\left(
K_{3\lambda }q^{\lambda }\right) ^{\dag }=q^{\lambda }K_{3\lambda }^{\dag }.
\label{s3}
\end{eqnarray}

To complete the remainder of the Sachs equation we need to evaluate the
quaternionic version of the scalar curvature, that is,%
\begin{equation}
\mathcal{R}\equiv -\frac{1}{2}[\tilde{q}^{\lambda }K_{\rho \lambda }q^{\rho
}+\tilde{q}^{\lambda }q^{\rho }K_{\rho \lambda }^{\dag }-K_{\rho \lambda
}^{\dag }\tilde{q}^{\rho }q^{\lambda }-\tilde{q}^{\rho }K_{\rho \lambda
}q^{\lambda }].
\end{equation}%
Consider%
\begin{equation}
\tilde{q}^{\lambda }K_{\rho \lambda }q^{\rho }=-\tilde{q}^{\rho }K_{\rho
\lambda }q^{\lambda }=\tilde{q}^{0}K_{\rho 0}q^{\rho }+\tilde{q}^{1}K_{\rho
1}q^{\rho }+\tilde{q}^{2}K_{\rho 2}q^{\rho }+\tilde{q}^{3}K_{\rho 3}q^{\rho
}.
\end{equation}%
Using Eqs. (\ref{s1}) - Eq. (\ref{s3}) we find%
\begin{eqnarray}
&&\tilde{q}^{0}K_{\rho 0}q^{\rho }+\tilde{q}^{1}K_{\rho 1}q^{\rho }+\tilde{q}%
^{2}K_{\rho 2}q^{\rho }+\tilde{q}^{3}K_{\rho 3}q^{\rho }  \notag \\
&=&e^{-2\lambda }[-\nu ^{\prime \prime }+\lambda ^{\prime }\nu ^{\prime
}-\nu ^{\prime 2}-\frac{2(\nu ^{\prime }-\lambda ^{\prime })}{r}+\frac{1}{%
r^{2}}(e^{2\lambda }-1)]
\end{eqnarray}%
Next, consider%
\begin{eqnarray}
\tilde{q}^{\lambda }q^{\rho }K_{\rho \lambda }^{\dag } &=&\left( K_{\rho
\lambda }q^{\rho }\tilde{q}^{\lambda }\right) ^{\dag },  \notag \\
K_{\rho \lambda }q^{\rho }\tilde{q}^{\lambda } &=&K_{\rho 0}q^{\rho }\tilde{q%
}^{0}+K_{\rho 1}q^{\rho }\tilde{q}^{1}+K_{\rho 2}q^{\rho }\tilde{q}%
^{2}+K_{\rho 3}q^{\rho }\tilde{q}^{3}  \notag \\
&=&-\frac{e^{\nu -2\lambda }}{2}[\nu ^{\prime \prime }-\lambda ^{\prime }\nu
^{\prime }+\nu ^{\prime 2}+\frac{2\nu ^{\prime }}{r}]\tilde{q}^{0}-\frac{%
e^{-\lambda }}{2}[\nu ^{\prime \prime }-\lambda ^{\prime }\nu ^{\prime }+\nu
^{\prime 2}-\frac{2\lambda ^{\prime }}{r}]\mathbf{\hat{x}\cdot \sigma }%
\tilde{q}^{1}  \notag \\
&&-\frac{e^{-2\lambda }}{2}[\nu ^{\prime }-\lambda ^{\prime }-\frac{%
(e^{2\lambda }-1)}{r}]\mathbf{\hat{\theta}\cdot \sigma }\tilde{q}^{2}-\frac{%
\sin \theta e^{-2\lambda }}{2}[\nu ^{\prime }-\lambda ^{\prime }-\frac{%
(e^{2\lambda }-1)}{r}]\mathbf{\hat{\phi}\cdot \sigma }\tilde{q}^{3}  \notag
\\
&=&e^{-2\lambda }[-\nu ^{\prime \prime }+\lambda ^{\prime }\nu ^{\prime
}-\nu ^{\prime 2}-\frac{2(\nu ^{\prime }-\lambda ^{\prime })}{r}+\frac{%
(e^{2\lambda }-1)}{r^{2}}]=\tilde{q}^{\lambda }q^{\rho }K_{\rho \lambda
}^{\dag }=\tilde{q}^{\lambda }K_{\rho \lambda }q^{\rho },  \notag \\
&&
\end{eqnarray}%
and similarly%
\begin{eqnarray}
K_{\rho \lambda }^{\dag }\tilde{q}^{\rho }q^{\lambda } &=&K_{\rho 0}^{\dag }%
\tilde{q}^{\rho }q^{0}+K_{\rho 1}^{\dag }\tilde{q}^{\rho }q^{1}+K_{\rho
2}^{\dag }\tilde{q}^{\rho }q^{2}+K_{\rho 3}^{\dag }\tilde{q}^{\rho }q^{3} 
\notag \\
&=&K_{\rho 0}\tilde{q}^{\rho }q^{0}+K_{\rho 1}^{\dag }\tilde{q}^{\rho
}q^{1}+K_{\rho 2}^{\dag }\tilde{q}^{\rho }q^{2}+K_{\rho 3}^{\dag }\tilde{q}%
^{\rho }q^{3}.
\end{eqnarray}%
Now%
\begin{eqnarray}
K_{\rho 0}^{\dag }\tilde{q}^{\rho } &=&K_{10}^{\dag }q^{1}+K_{20}^{\dag
}q^{2}+K_{30}^{\dag }q^{3}  \notag \\
&=&-\frac{e^{\nu -2\lambda }}{2}[\nu ^{\prime \prime }-\lambda ^{\prime }\nu
^{\prime }+\nu ^{\prime 2}+\frac{2\nu ^{\prime }}{r}],  \notag \\
K_{\rho 1}^{\dag }\tilde{q}^{\rho } &=&K_{01}^{\dag }\tilde{q}%
^{0}-K_{12}^{\dag }q^{2}-K_{13}^{\dag }q^{3}  \notag \\
&=&K_{10}q^{0}+K_{12}q^{2}+K_{13}q^{3}  \notag \\
&=&\frac{e^{-\lambda }}{2}[\nu ^{\prime \prime }-\lambda ^{\prime }\nu
^{\prime }+\nu ^{\prime 2}-\frac{2\lambda ^{\prime }}{r}]\mathbf{\hat{x}%
\cdot \sigma ,}  \notag \\
K_{\rho 2}^{\dag }\tilde{q}^{\rho } &=&K_{02}^{\dag }\tilde{q}%
^{0}-K_{21}^{\dag }q^{2}-K_{23}^{\dag }q^{3}  \notag \\
&=&K_{20}q^{0}+K_{21}q^{2}+K_{32}q^{3}  \notag \\
&=&\frac{e^{-2\lambda }}{2}[\nu ^{\prime }-\lambda ^{\prime }-\frac{1}{r}%
(e^{2\lambda }-1)]\mathbf{\hat{\theta}\cdot \sigma ,}  \notag \\
K_{\rho 3}^{\dag }\tilde{q}^{\rho } &=&K_{03}^{\dag }\tilde{q}%
^{0}-K_{31}^{\dag }q^{1}-K_{32}^{\dag }q^{2}  \notag \\
&=&K_{30}q^{0}+K_{31}q^{1}+K_{32}q^{2}  \notag \\
&=&\frac{\sin \theta e^{-2\lambda }}{2}[\nu ^{\prime }-\lambda ^{\prime }-%
\frac{1}{r}(e^{2\lambda }-1)]\mathbf{\hat{\phi}\cdot \sigma .}
\end{eqnarray}%
and so%
\begin{eqnarray}
K_{\rho \lambda }^{\dag }\tilde{q}^{\rho }q^{\lambda } &=&K_{\rho 0}^{\dag }%
\tilde{q}^{\rho }q^{0}+K_{\rho 1}^{\dag }\tilde{q}^{\rho }q^{1}+K_{\rho
2}^{\dag }\tilde{q}^{\rho }q^{2}+K_{\rho 3}^{\dag }\tilde{q}^{\rho }q^{3} 
\notag \\
&=&\frac{e^{-2\lambda }}{2}[\nu ^{\prime \prime }-\lambda ^{\prime }\nu
^{\prime }+\nu ^{\prime 2}+\frac{2\nu ^{\prime }}{r}]+\frac{e^{-2\lambda }}{2%
}[\nu ^{\prime \prime }-\lambda ^{\prime }\nu ^{\prime }+\nu ^{\prime 2}-%
\frac{2\lambda ^{\prime }}{r}]  \notag \\
&&+\frac{e^{-2\lambda }}{2r}[\nu ^{\prime }-\lambda ^{\prime }-\frac{1}{r}%
(e^{2\lambda }-1)]+\frac{e^{-2\lambda }}{2r}[\nu ^{\prime }-\lambda ^{\prime
}-\frac{1}{r}(e^{2\lambda }-1)]\mathbf{\hat{\phi}\cdot \sigma }  \notag \\
&=&e^{-2\lambda }[\nu ^{\prime \prime }-\lambda ^{\prime }\nu ^{\prime }+\nu
^{\prime 2}+\frac{2\nu ^{\prime }}{r}-\frac{2\lambda ^{\prime }}{r}-\frac{1}{%
r}(e^{2\lambda }-1)]=-\tilde{q}^{\lambda }q^{\rho }K_{\rho \lambda }^{\dag }.
\notag \\
&&
\end{eqnarray}%
Thus, we have%
\begin{eqnarray}
\mathcal{R} &\equiv &-\frac{1}{2}[\tilde{q}^{\lambda }K_{\rho \lambda
}q^{\rho }+\tilde{q}^{\lambda }q^{\rho }K_{\rho \lambda }^{\dag }-K_{\rho
\lambda }^{\dag }\tilde{q}^{\rho }q^{\lambda }-\tilde{q}^{\rho }K_{\rho
\lambda }q^{\lambda }]  \notag \\
&=&-\frac{1}{2}\{e^{-2\lambda }[-3\nu ^{\prime \prime }+3\lambda ^{\prime
}\nu ^{\prime }-3\nu ^{\prime 2}-\frac{6(\nu ^{\prime }-\lambda ^{\prime })}{%
r}+\frac{3}{r^{2}}(e^{2\lambda }-1)]  \notag \\
&&-e^{-2\lambda }[\nu ^{\prime \prime }-\lambda ^{\prime }\nu ^{\prime }+\nu
^{\prime 2}+\frac{2\nu ^{\prime }}{r}-\frac{2\lambda ^{\prime }}{r}-\frac{1}{%
r}(e^{2\lambda }-1)]\}  \notag \\
&=&-2[e^{-2\lambda }[-\nu ^{\prime \prime }+\lambda ^{\prime }\nu ^{\prime
}-\nu ^{\prime 2}-\frac{(2\nu ^{\prime }-2\lambda ^{\prime })}{r}+\frac{1}{%
r^{2}}(e^{2\lambda }-1)].
\end{eqnarray}%
Note this agrees with the scalar curvature obtained from Eq. (\ref{schw}) as
could be anticipated from Eq. ( \ref{sacre}). \ So, now we are in a position
to consider each term in the Sachs equation for $\rho =0,1,2,3.$ We obtain%
\begin{eqnarray}
&&\frac{1}{4}(K_{0\lambda }q^{\lambda }+q^{\lambda }K_{0\lambda }^{\dag })-%
\frac{1}{8}\mathcal{R}q_{0}=\frac{1}{8}(4K_{0\lambda }q^{\lambda }-\mathcal{R%
}q_{0})  \notag \\
&=&\frac{e^{\nu -2\lambda }}{8}[\frac{4\lambda ^{\prime }}{r}+\frac{2}{r^{2}}%
(e^{2\lambda }-1)]=0,  \notag \\
&&\frac{1}{4}(K_{1\lambda }q^{\lambda }+q^{\lambda }K_{1\lambda }^{\dag })-%
\frac{1}{8}\mathcal{R}q_{1}=\frac{1}{8}(4K_{1\lambda }q^{\lambda }-\mathcal{R%
}q_{1})  \notag \\
&=&\frac{e^{-\lambda }\mathbf{\hat{x}\cdot \sigma }}{8}[-\frac{4\nu ^{\prime
}}{r}+\frac{2}{r^{2}}(e^{2\lambda }-1)]=0\mathbf{,}  \notag \\
&&\frac{1}{4}(K_{2\lambda }q^{\lambda }+q^{\lambda }K_{2\lambda }^{\dag })-%
\frac{1}{8}\mathcal{R}q_{2}=\frac{1}{8}(4K_{2\lambda }q^{\lambda }-\mathcal{R%
}q_{2})  \notag \\
&=&\frac{re^{-2\lambda }}{8}\mathbf{\hat{\theta}\cdot \sigma \lbrack -}2\nu
^{\prime \prime }+2\lambda ^{\prime }\nu ^{\prime }-2\nu ^{\prime 2}-\frac{%
2(\nu ^{\prime }-\lambda ^{\prime })}{r}\mathbf{]=}0\mathbf{,}  \notag \\
&&\frac{1}{4}(K_{3\lambda }q^{\lambda }+q^{\lambda }K_{3\lambda }^{\dag })-%
\frac{1}{8}\mathcal{R}q_{3}=\frac{1}{8}(4K_{3\lambda }q^{\lambda }-\mathcal{R%
}q_{3})  \notag \\
&=&\frac{r\sin \theta e^{-2\lambda }}{8}\mathbf{\hat{\phi}\cdot \sigma
\lbrack \mathbf{-}}2\nu ^{\prime \prime }+2\lambda ^{\prime }\nu ^{\prime
}-2\nu ^{\prime 2}-\frac{2(\nu ^{\prime }-\lambda ^{\prime })}{r}\mathbf{]}=0%
\mathbf{.}  \label{sc4}
\end{eqnarray}%
The general case, in which each of the four quaternions $q^{\mu }(x)$ has
four components is a much more complicated set of 16 coupled highly
nonlinear equations. In the case of Eq. (\ref{sc4}) each of the four
quaternionic Sachs equations has only one component, not four. This
simplicity is not surprising since our quaternionic ansatz Eq. (\ref{qstz})
has a similar property, that is each of the quaternions has only one
component.\ \ In summary these four equations give the three independent
equations%
\begin{eqnarray}
\frac{2\lambda ^{\prime }}{r}+\frac{1}{r^{2}}(e^{2\lambda }-1)) &=&0,  \notag
\\
-\frac{2\nu ^{\prime }}{r}+\frac{1}{r^{2}}(e^{2\lambda }-1) &=&0,  \notag \\
\mathbf{\mathbf{-}}\nu ^{\prime \prime }+\lambda ^{\prime }\nu ^{\prime
}-\nu ^{\prime 2}-\frac{(\nu ^{\prime }-\lambda ^{\prime })}{r} &=&0.
\end{eqnarray}%
Subtracting the first two equations in the above set and substituting in the
last two we obtain%
\begin{eqnarray}
\nu ^{\prime }+\lambda ^{\prime } &=&0,  \notag \\
-2r\nu ^{\prime }e^{2\nu }+(1-e^{2\nu }) &=&0,  \notag \\
-\nu ^{\prime \prime }-2\nu ^{\prime 2}-\frac{2\nu ^{\prime }}{r} &=&0.
\end{eqnarray}%
\ 

Thus we get agreement between the Schwarzschild metric derived directly from
the standard Einstein equation and from Eq. (\ref{qstz}) together with the
Sachs equation (\ref{fac}). \ 

\section{\protect\bigskip Summary and Conclusion}

We have presented an in depth review of Sachs' early work \cite{s1} on his
quaternionic field equations of general relativity, culminating in the Sachs
equation and its quaternionic conjugate given in Eq. (\ref{fac}). \ Since
this equation is of the form of a four-vector quaternion, it is a 16
component equation with the correct number of components to solve uniquely,
given the appropriate boundary conditions, for the 16 components of the
basic quaternionic four vector $q^{\mu }(x).$ \ The new result we present in
this paper is an exact solution of this equation corresponding to the
static, spherically symmetric conditions that Schwarzschild used in his
derivation. \ The quaternionic four vector $q^{\mu }(x)$ that we found in
Eq. (\ref{qstz}) constructs the metric via Eq. (\ref{qm}) and using the
Sachs equation (\ref{fac}) leads to differential equations in terms of the
parametric functions $\nu (r)$ and $\lambda (r)$ that agree with those
obtained in Schwarzschild's treatment of the Einstein equation. \ The Sachs
equation (\ref{fac}) is, as mentioned in the introduction, a factorized
version of the Einstein equation. In \cite{s1}-\cite{s3}Sachs claims that 10
of the 16 equations can be brought to symmetric tensor form and therefore
identified with with gravity (Einstein's equations) and that 6 of the 16
equations can be brought to an antisymmetric tensor form and thus be
identifed with electromagnetism (Maxwell's equations). (See appendix B).\
However, in our view, solving the Sachs equation, which has the same number
of components as $q^{\mu }(x),$ is more direct and economical than solving
either of the above two factorized combinations.

\bigskip \appendix \makeatletter  \@addtoreset{equation}{section} %
\makeatother  \setcounter{equation}{0} \ \renewcommand{\theequation}{%
\Alph{section}.\arabic{equation}}

\section{The Spin Curvature Tensor and the Riemann Curvature Tensor}

Here we review the Sachs derivation \cite{s1}-\cite{s3},\cite{berg} of the
results given in Eq. (\ref{sacre}). \ Recall $g_{;\lambda }^{\mu \nu }=0=%
\left[ q^{\mu }(x)\tilde{q}_{\mu }(x)\right] _{;\lambda }$ implies $q^{\mu
}(x)_{;\lambda }=0$. \ Also, recall that in terms of its transformation
properties, the quaternion $q_{\mu }\sim (\eta \eta ^{\dag })_{\mu },$ so 
\begin{eqnarray}
0 &=&q_{\mu ;\rho ;\lambda }-q_{\mu ;\lambda ;\rho }=[(\eta _{;\rho ;\lambda
}-\eta _{;\lambda ;\rho })\eta ^{\dag }+\eta (\eta _{;\rho ;\lambda }^{\dag
}-\eta _{;\lambda ;\rho }^{\dag })]  \notag \\
&&+([q_{\mu ;\rho ;\lambda }]-[q_{\mu ;\lambda ;\rho }]),  \label{A1}
\end{eqnarray}%
in which the last term refers to difference between covariant derivatives of 
$q^{\mu }$ as a four vector only. \ That should be given in terms of the
Riemann curvature tensor by%
\begin{equation}
\lbrack q_{\mu ;\rho ;\lambda }]-[q_{\mu ;\lambda ;\rho }]=R_{\kappa \mu
\rho \lambda }q^{\kappa }.
\end{equation}%
\ Hence%
\begin{eqnarray}
&&[(\eta _{;\rho ;\lambda }-\eta _{;\lambda ;\rho })\eta ^{\dag }]_{\mu
}+[\eta (\eta _{;\rho ;\lambda }^{\dag }-\eta _{;\lambda ;\rho }^{\dag
})]_{\mu }  \notag \\
&=&-R_{\kappa \mu \rho \lambda }q^{\kappa }  \notag \\
&=&K_{\lambda \rho }[\eta \eta ^{\dag }]_{\mu }+[\eta \eta ^{\dag }]_{\mu
}K_{\lambda \rho }^{\dag }  \notag \\
&\rightarrow &K_{\rho \lambda }q_{\mu }+q_{\mu }K_{\rho \lambda }^{\dag
}=-R_{\kappa \mu \rho \lambda }q^{\kappa }  \label{41a}
\end{eqnarray}%
Similarly we find that 
\begin{equation}
K_{\rho \lambda }^{\dag }\tilde{q}_{\mu }+\tilde{q}_{\mu }K_{\rho \lambda
}=R_{\kappa \mu \rho \lambda }\tilde{q}^{\kappa }.  \label{41b}
\end{equation}%
Now%
\begin{eqnarray}
&&\tilde{q}_{\gamma }\ref{41a}-\ref{41b}q_{\gamma }  \notag \\
&=&\tilde{q}_{\gamma }K_{\rho \lambda }q_{\mu }+\tilde{q}_{\gamma }q_{\mu
}K_{\rho \lambda }^{\dag }-K_{\rho \lambda }^{\dag }\tilde{q}_{\mu
}q_{\gamma }-\tilde{q}_{\mu }K_{\rho \lambda }q_{\gamma }=-R_{\kappa \mu
\rho \lambda }(\tilde{q}_{\gamma }q^{\kappa }+\tilde{q}^{\kappa }q_{\gamma })
\notag \\
&=&-2\sigma _{0}R_{\kappa \mu \rho \lambda }\delta _{\gamma }^{\kappa },
\end{eqnarray}%
and so we have the following relation between the Riemann curvature tensor
and the spin curvature tensor,%
\begin{equation}
\sigma _{0}R_{\gamma \mu \rho \lambda }=\frac{1}{2}[\tilde{q}_{\mu }K_{\rho
\lambda }q_{\gamma }-\tilde{q}_{\gamma }K_{\rho \lambda }q_{\mu }-\tilde{q}%
_{\gamma }q_{\mu }K_{\rho \lambda }^{\dag }+K_{\rho \lambda }^{\dag }\tilde{q%
}_{\mu }q_{\gamma }].  \label{rr}
\end{equation}%
We distinguish between the right and left hand side by defining the
quaternionic Riemann curvature tensor%
\begin{equation}
\mathcal{R}_{\gamma \mu \rho \lambda }=\frac{1}{2}[\tilde{q}_{\mu }K_{\rho
\lambda }q_{\gamma }-\tilde{q}_{\gamma }K_{\rho \lambda }q_{\mu }-\tilde{q}%
_{\gamma }q_{\mu }K_{\rho \lambda }^{\dag }+K_{\rho \lambda }^{\dag }\tilde{q%
}_{\mu }q_{\gamma }].
\end{equation}%
Although formally the two fourth rank tensors $R_{\gamma \mu \rho \lambda }$
and $\mathcal{R}_{\gamma \mu \rho \lambda }~$should be equivalent based on
Eqs. (\ref{A1}-\ref{41a}), direct\ verification with use of just Eq. (\ref%
{qm}), (\ref{om}), (\ref{52}-\ref{o3}) and connections has not been
determined in published papers. \ The Ricci tensor is obtained by%
\begin{equation}
\sigma _{0}g^{\gamma \lambda }R_{\gamma \mu \rho \lambda }=\sigma _{0}R_{\mu
\rho }=\frac{1}{2}[\tilde{q}_{\mu }K_{\rho \lambda }q^{\lambda }-\tilde{q}%
^{\lambda }K_{\rho \lambda }q_{\mu }-\tilde{q}^{\lambda }q_{\mu }K_{\rho
\lambda }^{\dag }+K_{\rho \lambda }^{\dag }\tilde{q}_{\mu }q^{\lambda
}]\equiv \mathcal{R}_{\mu \rho },
\end{equation}%
and the curvature scalar by%
\begin{equation}
\sigma _{0}R=g^{\mu \rho }\sigma _{0}R_{\mu \rho }=\frac{1}{2}[\tilde{q}%
^{\rho }K_{\rho \lambda }q^{\lambda }-\tilde{q}^{\lambda }K_{\rho \lambda
}q^{\rho }-\tilde{q}^{\lambda }q^{\rho }K_{\rho \lambda }^{\dag }+K_{\rho
\lambda }^{\dag }\tilde{q}^{\rho }q^{\lambda }]\equiv \mathcal{R},
\end{equation}%
with $\mathcal{R}_{\mu \rho }$ called the quaternionic Ricci tensor and $%
\mathcal{R}$ the quaternionic scalar curvature.

\bigskip 

\section{\protect\bigskip The Relation Between the Sachs equations and the
Einstein and Maxwell Equations.}

The Einstein equations and the structure of the Maxwell equations can be
reproduced from the Sachs equations by a process that begins with
multiplying the first of Eq. (\ref{fac}) \ on the right by $\tilde{q}%
_{\gamma }$ and the second on the left by $q_{\gamma }$ giving

\begin{eqnarray}
\frac{1}{4}(K_{\rho \lambda }q^{\lambda }\tilde{q}_{\gamma }+q^{\lambda
}K_{\rho \lambda }^{\dag }\tilde{q}_{\gamma })-\frac{1}{8}q_{\rho }\tilde{q}%
_{\gamma }\mathcal{R} &=&k\mathcal{\tilde{F}}_{\rho }\tilde{q}_{\gamma }, 
\notag \\
\frac{1}{4}(-q_{\gamma }\tilde{q}^{\lambda }K_{\rho \lambda }-q_{\gamma
}K_{\rho \lambda }^{\dag }\tilde{q}^{\lambda })-\frac{1}{8}q_{\gamma }\tilde{%
q}_{\rho }\mathcal{R} &=&kq_{\gamma }\mathcal{\tilde{F}}_{\rho }.
\end{eqnarray}%
Adding and subtracting these produces

\begin{eqnarray}
\frac{1}{4}(K_{\rho \lambda }q^{\lambda }\tilde{q}_{\gamma }-q_{\gamma }%
\tilde{q}^{\lambda }K_{\rho \lambda }+q^{\lambda }K_{\rho \lambda }^{\dag }%
\tilde{q}_{\gamma }-q_{\gamma }K_{\rho \lambda }^{\dag }\tilde{q}^{\lambda
})-\frac{1}{8}(q_{\rho }\tilde{q}_{\gamma }+q_{\gamma }\tilde{q}_{\rho })%
\mathcal{R} &=&k(\mathcal{F}_{\rho }\tilde{q}_{\gamma }+q_{\gamma }\mathcal{%
\tilde{F}}_{\rho }),  \notag \\
\frac{1}{4}(K_{\rho \lambda }q^{\lambda }\tilde{q}_{\gamma }+q_{\gamma }%
\tilde{q}^{\lambda }K_{\rho \lambda }+q^{\lambda }K_{\rho \lambda }^{\dag }%
\tilde{q}_{\gamma }+q_{\gamma }K_{\rho \lambda }^{\dag }\tilde{q}^{\lambda
})-\frac{1}{8}(q_{\rho }\tilde{q}_{\gamma }-q_{\gamma }\tilde{q}_{\rho })%
\mathcal{R} &=&k(\mathcal{F}_{\rho }\tilde{q}_{\gamma }-q_{\gamma }\mathcal{%
\tilde{F}}_{\rho }).
\end{eqnarray}%
Taking the trace of both equations%
\begin{eqnarray}
Tr[K_{\rho \lambda }(q^{\lambda }\tilde{q}_{\gamma }-q_{\gamma }\tilde{q}%
^{\lambda })+K_{\rho \lambda }^{\dag }(\tilde{q}_{\gamma }q^{\lambda }-%
\tilde{q}^{\lambda }q_{\gamma })]-\frac{1}{2}Tr(q_{\rho }\tilde{q}_{\gamma
}+q_{\gamma }\tilde{q}_{\rho })\mathcal{R} &=&4kTr(\mathcal{F}_{\rho }\tilde{%
q}_{\gamma }+q_{\gamma }\mathcal{\tilde{F}}_{\rho }),  \notag \\
Tr[K_{\rho \lambda }(q^{\lambda }\tilde{q}_{\gamma }+q_{\gamma }\tilde{q}%
^{\lambda })+K_{\rho \lambda }^{\dag }(\tilde{q}_{\gamma }q^{\lambda }+%
\tilde{q}^{\lambda }q_{\gamma })]-\frac{1}{2}Tr(q_{\rho }\tilde{q}_{\gamma
}-q_{\gamma }\tilde{q}_{\rho })\mathcal{R} &=&4k(\mathcal{F}_{\rho }\tilde{q}%
_{\gamma }-q_{\gamma }\mathcal{\tilde{F}}_{\rho }).
\end{eqnarray}%
Using Eq. (\ref{qm}) leads to%
\begin{eqnarray}
Tr[K_{\rho \lambda }(q^{\lambda }\tilde{q}_{\gamma }-q_{\gamma }\tilde{q}%
^{\lambda })+K_{\rho \lambda }^{\dag }(\tilde{q}_{\gamma }q^{\lambda }-%
\tilde{q}^{\lambda }q_{\gamma })]-g_{\rho \gamma }Tr\mathcal{R} &=&4kTr(%
\mathcal{F}_{\rho }\tilde{q}_{\gamma }+q_{\gamma }\mathcal{\tilde{F}}_{\rho
}),  \notag \\
2Tr[K_{\rho \gamma }+K_{\rho \gamma }^{\dag }]-\frac{1}{2}Tr(q_{\rho }\tilde{%
q}_{\gamma }-q_{\gamma }\tilde{q}_{\rho })\mathcal{R} &=&4k(\mathcal{F}%
_{\rho }\tilde{q}_{\gamma }-q_{\gamma }\mathcal{\tilde{F}}_{\rho }).
\end{eqnarray}%
Using, from Eqs. (\ref{sacre})%
\begin{eqnarray*}
\mathcal{R}_{\mu \rho } &=&-\frac{1}{2}[\tilde{q}^{\lambda }K_{\rho \lambda
}q_{\mu }+\tilde{q}^{\lambda }q_{\mu }K_{\rho \lambda }^{\dag }-K_{\rho
\lambda }^{\dag }\tilde{q}_{\mu }q^{\lambda }-\tilde{q}_{\mu }K_{\rho
\lambda }q^{\lambda }], \\
Tr\mathcal{R}_{\gamma \rho } &=&-\frac{1}{2}Tr[K_{\rho \lambda }(q_{\gamma }%
\tilde{q}^{\lambda }-q^{\lambda }\tilde{q}_{\gamma })+K_{\rho \lambda
}^{\dag }(\tilde{q}^{\lambda }q_{\gamma }-\tilde{q}_{\gamma }q^{\lambda
})]=TrR_{\gamma \rho } \\
\mathcal{R} &\mathcal{=}&\sigma _{0}R,
\end{eqnarray*}%
and that $Tr(q_{\rho }\tilde{q}_{\gamma }-q_{\gamma }\tilde{q}_{\rho })=0,$
we obtain%
\begin{eqnarray}
R_{\rho \gamma }-\frac{1}{2}g_{\rho \gamma }R &=&kTr(\mathcal{F}_{\rho }%
\tilde{q}_{\gamma }+q_{\gamma }\mathcal{\tilde{F}}_{\rho }),  \notag \\
Tr[K_{\rho \gamma }+K_{\rho \gamma }^{\dag }] &=&2kTr(\mathcal{F}_{\rho }%
\tilde{q}_{\gamma }-q_{\gamma }\mathcal{\tilde{F}}_{\rho }).
\end{eqnarray}%
The first equation is equivalent to Einstein equation. What is the structure
of second equation? Define the antisymmetric tensor 
\begin{equation}
\mathcal{F}_{\rho \gamma }=Tr[K_{\rho \gamma }+K_{\rho \gamma }^{\dag }]=-%
\mathcal{F}_{\gamma \rho }
\end{equation}%
It has 6 independent components. \ It is also the curl of a four vector%
\begin{eqnarray}
\mathcal{F}_{\rho \gamma } &=&Tr[K_{\rho \gamma }+h.c.]=Tr[\Omega _{\rho
},_{\gamma }+\Omega _{\gamma }\Omega _{\rho }-\Omega _{\gamma },_{\rho
}-\Omega _{\rho }\Omega _{\gamma }+h.c.]  \notag \\
&=&Tr[\Omega _{\rho },_{\gamma }-\Omega _{\gamma },_{\rho }+h.c.]  \notag \\
&=&\mathcal{A}_{\rho },_{\gamma }-\mathcal{A}_{\gamma },_{\rho }
\end{eqnarray}%
Defining a current by 
\begin{equation}
\mathcal{F}_{~~~;\gamma }^{\rho \gamma }=2kTr(\mathcal{F}^{\rho }\tilde{q}%
^{\gamma }-q^{\gamma }\mathcal{\tilde{F}}^{\rho })_{;\gamma }\equiv j^{\rho
},
\end{equation}%
As a consequence of the antisymmetry we have that this current is
covariantly conserved%
\begin{equation*}
\mathcal{F}_{~~~;\gamma ;\rho }^{\rho \gamma }=0=j_{;\rho }^{\rho }=0.
\end{equation*}%
And, since it also the curl of a four vector 
\begin{equation}
\mathcal{F}_{\rho \gamma ;\lambda }+\mathcal{F}_{\gamma \lambda ;\rho }+%
\mathcal{F}_{\lambda \rho ;\gamma }=\mathcal{F}_{\rho \gamma ,\lambda }+%
\mathcal{F}_{\gamma \lambda ,\rho }+\mathcal{F}_{\lambda \rho ,\gamma }=0
\end{equation}%
These equations have the derived structure of the Maxwell equations
including the no magnetic charge and Faraday laws. \ However, the tensor $%
\mathcal{F}_{\rho \gamma }$ is not equivalent to the Faraday tensor since
from Eq. (\ref{trk}) we have that%
\begin{equation*}
\mathcal{F}_{\rho \gamma }=0.
\end{equation*}

\begin{acknowledgement}
The authors wish to express their special gratitude to Professor Mendel
Sachs \ for \ many useful and enlightening exchanges on his various works.
\end{acknowledgement}

\end{document}